\documentclass{aastex62}

\usepackage{amsfonts}
\usepackage{CJK}
\usepackage{graphicx}
\usepackage{subfigure}
\usepackage{amsmath,amsfonts,amssymb}
\usepackage{color}
\usepackage{booktabs}
\usepackage{multirow}
\usepackage{makecell}
\usepackage{hyperref}
\usepackage[normalem]{ulem}
\usepackage{CJKulem}
\def\iso#1{$^{#1}$}
\def\msun{M$_\odot$}

\def\nuc#1#2{\relax\ifmmode{}^{#1}{\protect\text{#2}}\else${}^{#1}$#2\fi}

\submitjournal{ApJL}

\shorttitle{The stellar $\beta$-decay rate of ${}^{134}$Cs}
\shortauthors{Li et al.}

\begin{document}

\title{The stellar $\beta$-decay rate of ${}^{134}$Cs and its impact on the Barium nucleosynthesis in the $s$ process}

\correspondingauthor{Kuo-Ang Li}
\email{lika@impcas.ac.cn}
\correspondingauthor{Chong Qi}
\email{chongq@kth.se}
\correspondingauthor{Maria Lugaro}
\email{maria.lugaro@csfk.org}

\author{Kuo-Ang Li}
\affil{CAS Key Laboratory of High Precision Nuclear Spectroscopy, Institute of Modern Physics, Chinese Academy of Sciences, Lanzhou 73000, People's Republic of China}
\affiliation{School of Nuclear Science and Technology, University of Chinese Academy of Sciences, Beijing 100049, People's Republic of China}
\author{Chong Qi}
\affiliation{Department of Physics, Royal Institute of Technology, Stockholm, Sweden}

\author{Maria Lugaro}
\affiliation{Konkoly Observatory, Research Centre for Astronomy and Earth Sciences, E\"otv\"os Lor\'and Research Network (ELKH), Konkoly Thege Mikl\'{o}s \'{u}t 15-17, H-1121 Budapest, Hungary}
\affiliation{ELTE E\"{o}tv\"{o}s Lor\'and University, Institute of Physics, Budapest 1117, P\'azm\'any P\'eter s\'et\'any 1/A, Hungary}
\affiliation{School of Physics and Astronomy, Monash University, VIC 3800, Australia}
\author{Andr{\'e}s Yag{\"u}e L{\'o}pez}
\affiliation{Konkoly Observatory, Research Centre for Astronomy and Earth Sciences, E\"otv\"os Lor\'and Research Network (ELKH), Konkoly Thege Mikl\'{o}s \'{u}t 15-17, H-1121 Budapest, Hungary}
\author{Amanda I. Karakas}
\affiliation{School of Physics and Astronomy, Monash University, VIC 3800, Australia}
\affiliation{ARC Centre of Excellence for All Sky Astrophysics in 3 Dimensions (ASTRO 3D), Canberra, ACT 2611, Australia}
\author{Jacqueline den Hartogh}
\affiliation{Konkoly Observatory, Research Centre for Astronomy and Earth Sciences, E\"otv\"os Lor\'and Research Network (ELKH), Konkoly Thege Mikl\'{o}s \'{u}t 15-17, H-1121 Budapest, Hungary}
\author{Bing-Shui Gao}
\affil{CAS Key Laboratory of High Precision Nuclear Spectroscopy, Institute of Modern Physics, Chinese Academy of Sciences, Lanzhou 73000, People's Republic of China}
\affiliation{School of Nuclear Science and Technology, University of Chinese Academy of Sciences, Beijing 100049, People's Republic of China}
\author{Xiao-Dong Tang}
\affil{CAS Key Laboratory of High Precision Nuclear Spectroscopy, Institute of Modern Physics, Chinese Academy of Sciences, Lanzhou 73000, People's Republic of China}
\affiliation{School of Nuclear Science and Technology, University of Chinese Academy of Sciences, Beijing 100049, People's Republic of China}
\affiliation{Joint department for nuclear physics, Institute of Modern Physics and Lanzhou University, Lanzhou, China, 730000}

\begin{abstract}
We have calculated the stellar $\beta$-decay rate of the important s-process branching point ${}^{134}$Cs based on the state of the art shell model calculations.  At typical $s$-process temperatures ($T\sim$ 0.2-0.3 GK), our new rate is one order of magnitude lower than the widely-used rate from Takahashi and Yokoi (hereafter TY87). The impact on the nucleosynthesis in AGB stars is investigated with various masses and metallicities. Our new decay rate leads to an overall decrease in the ${}^{134}$Ba/${}^{136}$Ba ratio, and well explains the measured ratio in meteorities without introducing the $i$ process. We also derive the elapsed time from the last AGB nucleosynthetic event that polluted the early Solar System to be $>$28 Myr based on the ${}^{135}$Cs/${}^{133}$Cs ratio, which is consistent with the elapsed times derived from ${}^{107}$Pd and ${}^{182}$Hf. The $s$-process abundance sum of ${}^{135}$Ba and ${}^{135}$Cs is found to increase, resulting in a smaller $r$-process contribution of ${}^{135}$Ba in the Solar System.

\end{abstract}


\section{Introduction} \label{sec:intro}
\begin{figure}[htbp]
  \centering
  \includegraphics[width=0.46\textwidth]{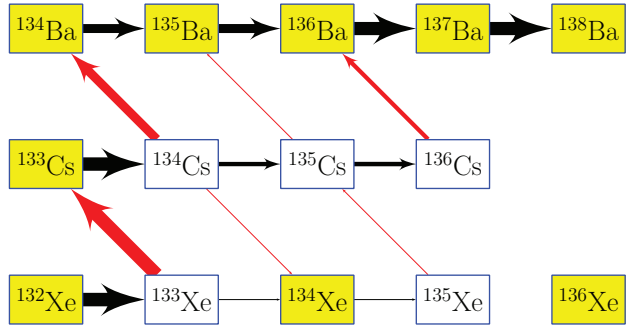}
  \caption{The s-process path at the vicinity of \nuc{134}{Cs} at $T =$ 0.35 GK($kT$ = 30 keV) generated by the NUCNET code\citep{nucnet}. The red lines present $\beta$-decay, while the black lines present the neutron capture. Line width indicates the flow current in linear scale. The stable nuclei are marked with yellow color.}
\label{branching}
\end{figure}

The \nuc{134}{Cs} stellar $\beta-$decay rate is crucial to understand the origin of the Ba isotopes.
The competition between the neutron capture and the $\beta$-decay of \nuc{134}{Cs}
determines the relative abundances of \iso{134}Ba and \iso{136}Ba, two pure $s$-process isotopes shielded by their stable
xenon isobars, and influences the $s$-process nucleosynthesis of \iso{135}Ba and \iso{135}Cs. The astrophysical sites for the production of these two Ba isotopes in the
Galaxy are asymptotic giant branch (AGB) stars with initial masses predominantly in the range 2-4 \msun. The main neutron source is
the \iso{13}C($\alpha$,n)\iso{16}O reaction activated in a thin region of the He-rich shell at $T\sim0.1$GK \citep{gallino98,travaglio99,karakas14dawes}.The \iso{22}Ne($\alpha$,n)\iso{25}Mg reaction is also activated during the thermal instabilities associated with He
burning in these stars at $T\sim0.3$GK. It provides a peak neutron density
orders of magnitude higher than that generated by the \iso{13}C source:
up to $10^{12-13}$ cm$^{-3}$ ($^{22}$Ne source) compared to $10^{7-8}$
cm$^{-3}$ ($^{13}$C source). As a result, the branching point \iso{134}Cs
is much more efficiently activated in the neutron flux provided by the \iso{22}Ne than by
the \iso{13}C neutron source\citep{lugaro03,fishlock14,bisterzo15}. A typical $s$-process path is shown in Figure~\ref{branching}.

The rich variety of observations of the Ba and Cs isotopes in the Universe, which related to this branching, offers a number of opportunities to probe the nucleosynthesis at various sites: (i)It was found that two nominally mainstream SiC grains with strongly
negative $\delta$(\iso{134}Ba/\iso{136}Ba) values could not be explained by any of the
current AGB model calculations, suggesting that such negative values
were consistent with the intermediate neutron capture process ($i$-process), whose neutron density is much higher than that of s-process, during the post-AGB stage\citep{liu14}. (ii)\iso{135}Cs($T_{1/2}$=2.3 Myr) is a long-lived isotope produced by AGB stars. It
can be used to derive the elapsed time from the last AGB nucleosynthetic event that polluted the pre-solar system matter to the birth of the Sun\citep{brennecka17}. (iii)\iso{135}Ba is produced in the $s-$ and $r-$processes. A reliable prediction of
the abundance sum of \iso{135}Ba and its progenitor nucleus, \iso{135}Cs,
in AGB stars is important to determine the $r$-process contribution in the Solar System.

A reliable $\beta$-decay rate of \iso{134}Cs is essential for describing quantitatively the branching ratio and interpret correctly these nucleosynthetic outcomes.
In stellar environments, the low-lying states of \iso{134}Cs are
thermally populated and contribute to the total $\beta$-decay rates.
Takahashi and Yokoi calculated the beta decay rate of \iso{134}Cs
based on the empirical log$ft$ values extracted from the outdated nuclear structure data in the 1980s~\citep{Takahashi87}.
By artificially varying the transition strengths, the uncertainty of the \nuc{134}{Cs} $\beta$-decay has been estimated to be of one order of magnitude~\citep{Goriely1999}. As the neutron capture cross sections on \iso{134}Cs and \iso{134,136}Ba
are determined with errors better than 10\%~\citep{patronis2004,KADoNiS},
the uncertainty of the $\beta$-decay rate is the major limitation in the studies of the $s$ process.

In this paper, we present a new stellar
$\beta$-decay rate of \nuc{134}{Cs} deduced from the Gamow-Teller (GT)
transition strength [$B$(GT)] obtained from large-scale shell-model
calculations. With the new $\beta$-decay rate, we investigate its impact on
the nucleosynthesis using AGB stellar models covering various masses and
metallicities.

\section{Calculation of the $\beta-$decay rate of \nuc{134}{Cs} in the stellar environment}
\begin{figure}[htbp]
  \centering
  \includegraphics[width=0.40\textwidth]{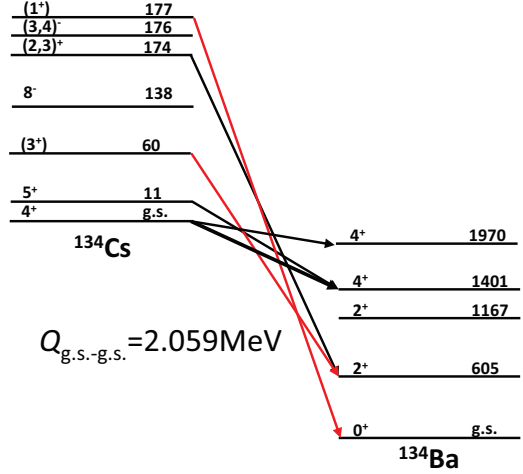}
  \caption{The stellar $\beta$-decay scheme of \nuc{134}{Cs}. The weak $\beta^{+}$/EC channel is not shown. Only allowed transitions of excited states below 200keV are listed. The important transitions to low-lying states of \nuc{134}{Ba} are highlighted with red lines. The tentative $J^{\pi}$ assignments are shown with parentheses. }
\label{cs134decay}
\end{figure}
The terrestrial decay rate of \nuc{134}{Cs} has been well studied in the laboratory with a half-life of 2.0652(4) years. It is dominated by the
$\beta^{-}$ decay. Only a tiny fraction of the decay channels,
$3\times10^{-6}$, is $\beta^{+}$/electron-capture(EC) decay,
which is negligible. In stellar environments, $\beta$-decays from thermally populated excited
states could make a significant contribution to total decay rates.
However, even at the $s$ process temperatures, $\beta^+$/EC decays
contribute less than 1\% of the total decay rate according to TY87. Due to its minor contribution, in the present work we focus on the $\beta^{-}$-decay channel and adopt TY87 rate for the $\beta^{+}-$/EC channel.

Fig~\ref{cs134decay} shows the stellar $\beta$-decay scheme of \nuc{134}{Cs}.
The terrestrial \nuc{134}{Cs} $\beta$-decay is dominated by
transitions from the \nuc{134}{Cs} $4^+$ ground state to the two $4^+$
states of  \nuc{134}{Ba} with excitation energies of  1401 and
1970 keV, respectively. Due to the small decay energies (658 and 89 keV),
the terrestrial decay rate is relatively slow.
However, with the thermal population of the excited states, the stellar $\beta$-decay rate could be significantly
enhanced due to contributions from the two transitions, ($3^+$, 60 keV)$\rightarrow$($2^+$, 605 keV) and
($1^+$, 177 keV)$\rightarrow$($0^+$, g.s.), in the $s$-process environment,
due to the combination with high population possibility, high transition strength and the large decay phase space.

In the TY87 rate, the unknown transition strengths of the excited states were
estimated from the analogous transitions in the existing experimental data, i.e.,
the transitions from the terrestrial $\beta$-decay of neighbouring nuclei.
Nowadays with the progress in computational nuclear physics, the transition
strength can be calculated more precisely by using
the large-scale nuclear shell model.

We have applied the large-scale shell model (hereafter SM) to calculate the transition
strengths of the excited states and updated the stellar $\beta$-decay rate of
\nuc{134}{Cs}. The large-scale shell model is a full configuration
interaction approach that takes into account all possible couplings and
correlations among valence nucleons within one or a few major shells. We assume $^{100}$Sn as the inert core  and  consider the
neutron and proton orbitals between the shell closures $N$ (and $Z$) $=50$
and 82, comprising $0g_{7/2}$, $1d_{5/2}$, $1d_{3/2}$, $2s_{1/2}$ and
$0h_{11/2}$ orbitals. The starting point of our calculation is the realistic CD-Bonn
nucleon-nucleon potential~\citep{machleidt01}. The interaction was renormalized using the
perturbative G-matrix approach to take into account the core-polarization
effects~\citep{jensen95}. The $T=1$ component of the monopole interaction was optimized by fitting
the low-lying states of all Sn isotopes between $^{101}$Sn and $^{132}$Sn
(see Ref.~\citet{chong12} for details).
Our calculations in the present work reproduced the spectrum of \iso{134}Ba and all the lying states of \iso{134}Cs in Fig.~\ref{cs134decay} with recommended spin-parity assignments within ${\pm}200$ keV, though the orders can be slightly different. It is, however, not a major issue for the present work because the levels and their corresponding wave functions are identified based on their J$^\pi$ values. The wave functions of \iso{134}Cs and \iso{134}Ba are calculated to be dominated by the coupling of protons in nearly degenerate $g_{7/2}d_{5/2}$ orbitals and neutrons in the $s_{1/2}d_{3/2}h_{11/2}$ orbitals. Those are consistent with the low-lying spectra of neighbouring odd-A \iso{133,135}Cs and \iso{133,135}Ba. As a result, the GT strength mostly come from the transition ${\pi}d_{5/2}$ to ${\nu}d_{3/2}$.

\begin{figure}[htbp]
  \centering
  \includegraphics[width=0.41\textwidth]{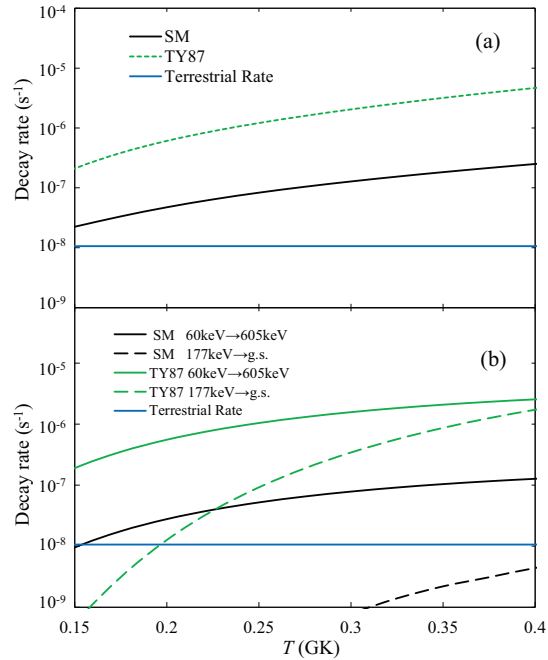}
  \caption{\iso{134}Cs stellar $\beta^{-}-$decay rate of TY87 and the present work obtained with the shell model (SM)(a), and the decay rates of individual important transitions (b). The terrestrial rate is also presented. }
  \label{ratio_transition}
\end{figure}

With the updated log$ft$ values calculated from the shell model,
we obtained a new stellar $\beta^{-}$-decay rate of \iso{134}Cs which is shown in Fig~\ref{ratio_transition}(a) along with TY87 rate.
The new rate is about one order of magnitude slower than TY87 at the $s$-process temperatures.

The individual contributions of the major transitions in the total stellar $\beta^{-}$-decay rate of \iso{134}Cs is presented in Fig~\ref{ratio_transition}(b). The ($3^+$, 60 keV)$\rightarrow$($2^+$, 605 keV) transition (log$ft=$ 7.8(SM) vs. 6.5(TY87)) dominates, and the transition ($1^+$, 177 keV)$\rightarrow$($0^+$, g.s.) contributes only a minor fraction (log$ft=$ 7.6(SM) vs. 5.5(TY87)).
To estimate the uncertainty of the transition strength calculated by the shell model calculation, we have done several calculations by slightly varying nucleon excitations from  $g_{7/2}d_{5/2}$ to $s_{1/2}d_{3/2}h_{11/2}$ orbitals and the single-particle energies of the neutron $d_{3/2}$ orbital. In those calculations the calculated GT strength for the mostly relevant $3^+$ to $2^+$ decay vary mostly within a factor of two. As the most important transition ($3^+$, 60 keV)$\rightarrow$($2^+$, 605 keV) is between two short lived excited states, our approach is the best way at present to obtain the strength.

One major source of the uncertainty of the \nuc{134}{Cs} decay rate comes from the spin-parity ($J^{\pi}$) assignments of the
excited states. In our shell model calculations and TY87, the level scheme is taken from ~\citet{NNDC}. However, due to the limitations
of experimental data, the $J^{\pi}$ assignments of some states are still debated. $J^{\pi}$ = $3^+$ is suggested for the 60-keV state
in which case the transition to the ($2^+$, 605keV) state in $^{134}$Ba is
allowed. However, other $J^{\pi}$ values such as $4^+$ could not be excluded~\citep{NNDC,cs134-level} resulting in
a forbidden transition with a significant reduction in the total decay rate.
A similar situation also exists for the 177-keV state. The transition from this state to the \nuc{134}{Ba} ground state will be of
forbidden nature if its $J^\pi$ is not $1^+$.
Such ambiguity could be eliminated by further
experiment such as the in-beam $\gamma$ spectroscopy.

\section{Impact of new \nuc{134}{Cs} $\beta-$decay rate on nucleosynthesis}
To test the impact of the new decay rate presented
here on the production of the Ba isotopes in the Galaxy, we have run
several selected models of AGB nucleosynthesis for masses in the range 2
to 4.5 \msun and metallicities from 1/140 to twice solar metallicity
\citep{lugaro12,karakas14,karakas16,karakas18}. A solar metallicity of
0.014 \citep{asplund09} is adopted in the present work. As described in more detail in, e.g., \citet{karakas16}, our computational method is based on the
post-processing nucleosynthesis code developed by \citet{cannon93},
where the changes in the abundances due to nucleosynthesis and mixing
within convective regions are solved simultaneously. This means that we
can treat the temperature dependence of the decay rate of \iso{134}Cs
together with mixing during the thermal instability where the \iso{22}Ne
neutron source is activated. In any case, our results appear generally robust
considering that for our  \iso{134}Ba/\iso{136}Ba ratios are
remarkably similar (within 10\%) to those produced by the FRUITY models
\citep{cristallo09,cristallo11} for the same stellar masses and
metallicities. The FRUITY models were calculated by interpolating the TY87 decay rates over the temperature of interest (i.e., not using
an average value of the rate), but differently to our computations they
solved separately the effects of
nucleosynthesis and of the mixing.

\begin{deluxetable*}{cccccc}
\tablewidth{0pc}
\tablecaption{Stellar surface abundances by number (in units of $10^{-11}$) and isotopic ratios for selected Ba and Cs isotopes at the end of the evolution for our set of AGB models. Values in roman font are calculated using the $^{134}$Cs decay rate by \citet{Takahashi87}, while values given in italics are calculated using the decay rate from the present study, with the change indicated in brackets. \label{tab:AGBmodels}}
\tablehead{
\colhead{Mass (M$_\odot$)} & \colhead{$^{134}$Ba} & \colhead{$^{135}$Ba} & \colhead{$^{135}$Ba+$^{135}$Cs} & \colhead{($^{134}$Ba/$^{136}$Ba)/$\odot$} & \colhead{$^{135}$Cs} \\
\hline
\multicolumn{6}{c}{Z = 0.03}
}
\startdata
\hline
2.5 & 10.2, {\it 9.53 (-7\%)} & 4.86, {\it 4.67 (-4\%)} & 5.15, {\it 6.00 (+16\%)} & 1.38, {\it 1.25} & 0.29, {\it 1.33 ($\times$4.6)}  \\
2.75 & 10.6, {\it 9.53 (-10\%)} & 5.06, {\it 4.74 (-6\%)}  & 5.55, {\it 6.63 (+20\%)} & 1.33, {\it 1.17} & 0.49, {\it 1.89 ($\times$3.8)} \\
3 & 9.29, {\it 7.93 (-15\%)} & 4.57, {\it 4.13 (-10\%)} & 5.21, {\it 6.46 (+24\%)} & 1.28, {\it 1.08} & 0.64, {\it 2.33 ($\times$3.6)} \\
3.5 & 4.67, {\it 3.75 (-20\%)} & 2.87, {\it 2.58 (-10\%)}  & 3.47, {\it 4.32 (+25\%)} & 1.22, {\it 0.97} & 0.60, {\it 1.75 ($\times$2.9)} \\
4 & 3.48, {\it 2.46 (-29\%)} & 2.40, {\it 2.00 (-17\%)} & 2.99, {\it 3.73 (+25\%)} & 1.15, {\it 0.82}  & 0.71, {\it 1.79 ($\times$2.5)} \\
4.5 & 1.07, {\it 0.88 (-18\%)} & 1.59, {\it 1.51 (-5\%)} & 1.73, {\it 1.93 (+11\%)} & 1.12, {\it 0.93} & 0.17, {\it 0.42 ($\times$2.4)} \\
\hline
\multicolumn{6}{c}{Z = 0.014}\\
\hline
2 & 10.5, {\it 9.20 (-12\%)} & 4.43, {\it 4.01 (-10\%)} & 4.97, {\it 6.19 (+25\%)} & 1.28, {\it 1.12} & 0.54, {\it 2.18 ($\times$4.0)} \\
3 & 14.0, {\it 10.6 (-24\%)} & 5.67, {\it 4.65 (-18\%)} & 7.86, {\it 10.7 (+37\%)} & 1.18, {\it 0.90} & 2.19, {\it 6.10 ($\times$2.8)} \\
3.5 & 7.84, {\it 5.50 (-30\%)} & 3.34, {\it 2.59 (-22\%)} & 5.52, {\it 7.56 (+37\%)} & 1.16, {\it 0.82} & 2.18, {\it 4.97 ($\times$2.3)} \\
4 & 6.93, {\it 4.18 (-40\%)} & 2.95, {\it 1.83 (-38\%)} & 6.97, {\it 9.37 (+34\%)} & 1.16, {\it 0.71}  & 4.01, {\it 7.54 ($\times$1.9)} \\
4.5 & 2.32, {\it 1.48 (-36\%)} & 1.40, {\it 1.04 (-25\%)} & 2.72, {\it 3.49 (+29\%)}  & 1.11, {\it 0.71} & 1.32, {\it 2.45 ($\times$1.8)}  \\
\hline
\multicolumn{6}{c}{Z = 0.007}\\
\hline
2.5 & 14.8, {\it 11.0 (-26\%)} & 5.77, {\it 4.64 (-35\%)} & 8.50, {\it 12.0 (+41\%)} & 1.08, {\it 0.80}  & 2.72, {\it 7.32 ($\times$2.7)} \\
3 & 15.8, {\it 10.3 (-20\%)} & 6.01, {\it 4.48 (-25\%)} & 11.8, {\it 16.6 (+41\%)}  & 1.05, {\it 0.69} & 5.78, {\it 12.1 ($\times$2.1)}  \\
\hline
\multicolumn{6}{c}{Z = 0.0028}\\
\hline
3 & 4.92, {\it 2.84 (-42\%)} & 1.93, {\it 1.23 (-36\%)} & 5.67, {\it 7.45 (+31\%)}  & 1.04, {\it 0.61} & 3.74, {\it 6.22 ($\times$1.7)}  \\
\hline
\multicolumn{6}{c}{Z = 0.001}\\
\hline
2.5 & 1.90, {\it 1.07 (-36\%)} & 0.75, {\it 0.47 (-37\%)} & 2.01, {\it 2.70 (+34\%)} & 1.06, {\it 0.61}  & 1.26, {\it 2.23 ($\times$1.8)} \\
\hline
\multicolumn{6}{c}{Z = 0.0001}\\
\hline
2.5 & 0.19, {\it 0.11 (-42\%)} & 0.08, {\it 0.05 (-37\%)} & 0.26, {\it 0.33 (+28\%)} & 0.94, {\it 0.17}  & 0.18, {\it 0.28 ($\times$1.6)} \\
\enddata
\end{deluxetable*}

The results of our models are presented in Table~\ref{tab:AGBmodels}. Because
the new decay rate is typically lower than the TY87 rate, e.g., roughly by a factor of ten at 270 MK (23
keV), the neutron capture of \iso{134}Cs is more effectively activated in the models
calculated with the new rate. As the \iso{22}Ne reaction is dependent on
the temperature in the He shell, which in turn increases with increasing
stellar mass and decreasing metallicity, the significance of the impact of
the new rate also increases with increasing stellar mass and decreasing
metallicity. In summary, we obtained a significant decrease in the abundance of
\iso{134}Ba and \iso{135}Ba by up to roughly 40\%, a strong increase in \iso{135}Cs (by roughly a factor 2 to 4), and no significant changes in the \iso{136}Ba abundance,
which indicates that the branching path rejoins the standard s-process path at this isotope, in other words, there is
no further effect of the branching chain through the Cs isotopes on \iso{136}Ba and \iso{137}Ba.
Interestingly, since \iso{135}Cs eventually decays into \iso{135}Ba, the final s-process \iso{135}Ba abundance, i.e., the abundance sum of \iso{135}Ba and \iso{135}Cs, increases by up to 40\%.

\begin{figure*}
\center
\includegraphics[width=7in]{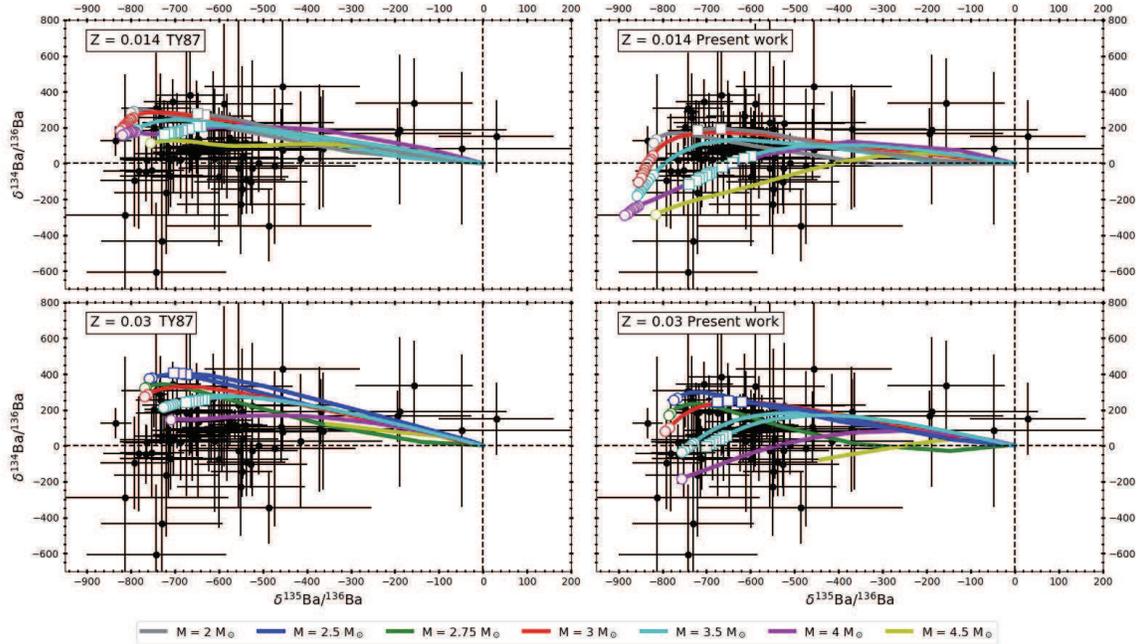}

\caption{Comparison between data for single stardust SiC grains
\citep[black dots with 2$\sigma$ error bars, from][]{liu14} and AGB
model predictions of different masses, metallicities, and decay rates of
\iso{134}Cs, as indicated by the labels, where Z = 0.014 is the solar
metallicity, TY87 is the rate from \citet{Takahashi87} and ``Present work''
refers to the new rate presented in this paper. The data are plotted
using the standard $\delta$ notation, where
$\delta$\iso{134,135}Ba/\iso{136}Ba represents the permil variation of the
ratio with respect to solar. The model predictions start at solar
composition ($\delta=0$) and follow the coloured lines until the envelope becomes C-rich (the condition for the formation of SiC) and open symbols
are plotted. This means that the SiC data points should be compared
exclusively to the open symbols. Circles represent models where
the extent in mass $M_{\rm PMZ}$ of the mixing of protons leading to the formation
of the \iso{13}C neutron source was taken to be the standard reported in
Table~1 of \citet{lugaro18a}. Squares represents models where $M_{\rm PMZ}$ was reduced by a factor of 10 in the Z=0.014 models and
by a factor of 2 in the Z=0.03 models. \label{fig:grains}}
\end{figure*}

The decrease in \iso{134}Ba results in an overall decrease of the
\iso{134}Ba/\iso{136}Ba isotopic ratio, with the geometric average of all the models shown in Table~\ref{tab:AGBmodels}
changing from 16\% higher than solar
using the old rate, to 16\% lower than solar using the new rate. However, a
proper comparison with the solar abundances can only be accomplished by
running full galactic chemical evolution models.

For the final $s$-process \iso{135}Ba abundance, the geometric average of all the models shown in Table~\ref{tab:AGBmodels} is higher by 30\% when using the new rate, relatively to using the TY87 rate, although a full galactic chemical evolution calculation is needed to derive the accurate number. Such variation results in a small decrease of roughly 12\% in the $r$-process residual abundance of \iso{135}Ba. When added to the present error bars of roughly 9\%, based on uncertainties on the neutron-capture cross section and the solar abundance~\citep{arlandini99}, the value of the $r$-process residual abundance of \iso{135}Ba may be up to 20\% lower than used usually. Considering, for example, Figure 6 of \citet{vockenhuber07}, this decrease would move the $r$-process residual abundance of \iso{135}Ba to just below that of \iso{137}Ba, while one would expect \iso{135}Ba to be above \iso{137}Ba. A full analysis of the branching points along the Cs isotopes leading also to the production of \iso{137}Ba is needed to evaluate and resolve this possible tension. If the second $r$-process peak (at $A\sim130$) results in being reshaped and potentially narrowed by the lower \iso{135}Ba residual derived using our new rate, this will have implications on our understating of the fission recycling during the $r$ process~\citep{kajino19,mumpower18}.

In Figure~\ref{fig:grains} we present the comparison between our model
predictions and the composition of single stardust SiC grains measured
with Resonant Ionisation Mass Spectrometry \citep[RIMS, see][and
references therein]{liu14}. We focus in this figure on the models with solar and higher-than-solar metallicity, as these has been reported
previously to provide the best match to the SiC data
\citep[e.g.,]{lugaro18a,liu18}. They are the same models used by
\citet{lugaro18a} and more details on their structural features can be found in
Table~1 of that paper. As discussed above, the models calculated with the
new \iso{134}Cs decay rate result in lower \iso{134}Ba/\iso{136}Ba ratios,
which, when compared to the grain data, provide a general better fit to
the observed spread. The new models reach down to cover, within the error
bars, the two grains most strongly depleted in \iso{134}Ba which were
previously unexplained and potentially attributed to the operation of the
intermediate neutron capture ($i$ process) in post-AGB stars
\citep{liu14}, suggesting that the two grains could still be explained by the $s$ process alone. Due to its being already significantly lower than solar, the predicted lines for the \iso{135}Ba/\iso{136}Ba ratio are only marginally shifted to the left due to the new rate.
The grain data indicates a spread of \iso{135}Ba/\iso{136}Ba ratio with values towards the solar value. This spread can
be predicted when changing the free parameter in the models which
represents the extent in mass of the region of partial mixing ($M_{\rm PMZ}$) leading to
the formation of the \iso{13}C neutron source \citep{karakas16,lugaro18a}. In the case of
the solar metallicity models we needed to divide $M_{\rm PMZ}$ by a factor
of 10, in the case of the models of twice-solar metallicity it was
enough to reduce this mass by a factor of two to derive a better match
to the measured \iso{135}Ba/\iso{136}Ba ratios. The difference is due to
the models of lower metallicity being more efficient in
producing the $s$-process isotopes, such as \iso{136}Ba, because of the lower
number of Fe seeds \citep{clayton88}. Finally we note that for this
comparison we considered the \iso{135}Ba abundance alone, without the
contribution of \iso{135}Cs. This assumes that SiC condenses in AGB
envelopes within a few Myr, i.e., before a significant amount of \iso{135}Cs
has decayed into \iso{135}Ba, and that Cs does not condense into SiC, since
it is a relatively volatile element. Our models support this assumptions,
which was already discussed in \citet{lugaro03} (see their Figure~16), and even more
strongly when considering the predictions obtained using the new decay rate
of \iso{134}Cs.

While our new rate has direct and indirect implications on the $s$- and $r$-process, respectively, it is not expected to lead to significant modification of the Ba isotopic abundances produced by the $i$ process, with neutron densities of the order of $10^{13}-10^{15}$ cm$^{-3}$, in-between those of the $s$ and $r$ processes.  This is because during the $i$ process the Ba isotope most produced is \iso{135}Ba, resulting from the decay of the neutron magic unstable \iso{135}I (see, e.g., Figure~1 of \citet{hampel16}). The behavior of the Cs branching points represents a less significant effect relative to this major feature of the $i$ process.

There are also potential implications of our results for the early Solar System.
Long-lived radioactive isotopes
have been found to be present at the time of the formation of the first solids at
the birth of the Sun, however, for \iso{135}Cs/\iso{133}Cs ratio only an upper limit of $2.8 \times 10^{-6}$ \citep{brennecka17} is
available. This means that from this upper limit we can only derive a lower
limit for the time elapsed from the last AGB nucleosynthetic event that
polluted the pre-solar system matter to the birth of the Sun \citep{lugaro14b}.
For consistency with \citep{lugaro14b,lugaro18b} we consider the 3 \msun,
Z=0.014 model as typical, and calculated the time that elapsed from the last
AGB events using Eq.~15 of \citet{lugaro18b} with the Galactic chemical
evolution parameter K=2.3 \citep{cote19a} and with the time interval between
polluting events of 50 Myr to ensure that we are in the regime where it is
only one event that contributed to the early Solar System abundances of the
long-lived isotopes produced by AGB stars \citep{cote19b}. We derive elapsed
times of the order of 32-38 Myr from \iso{107}Pd and \iso{182}Hf. For
\iso{135}Cs we derive an elapsed time $>$24 Myr and $>$28 Myr when
considering the \iso{135}Cs/\iso{133}Cs ratios predicted using the \iso{134}Cs
decay rate by TY87 and the present work, respectively. In both cases
agreement is found with the other two long-lived isotopes.

Finally, we examined the impact of the new decay rate on the ratio $f_{\rm
odd}$ between the abundances of the odd-A and even-A isotopes of Ba. In
most of our models $f_{\rm odd}$ is very close to traditional $s$-process
reported value of $f_{\rm odd}$ = 0.11 \citep{arlandini99}. The changes are
at most a increase of 15\% when considering the new decay
rate.

\section{Summary}
The stellar $\beta-$decay rate of the branching point \nuc{134}{Cs} is crucial to the understanding of $s$-process nucleosynthesis at the vicinity of \iso{134-136}Ba. Large-scale shell-model
calculations were performed to determine the GT strengths of the important
transitions from the \nuc{134}{Cs} excited states to the excited and ground
states of \nuc{134}{Ba}.
At typical $s$-process temperatures, the new decay rate from our calculation is significantly lower than
the TY87 rate estimated with empirical log$ft$ values.
With the new $\beta-$decay rate, we preformed nucleosynthesis
calculations using AGB stellar models with various masses and metallicities. Our
result shows an overall decrease in the $s$-only \iso{134}Ba/\iso{136}Ba
isotopic ratio and well explains the \iso{134}Ba/\iso{136}Ba isotopic ratios in meteorites without introducing the $i$ process at the post-AGB phase.
We also derive the elapsed time from the last AGB nucleosynthetic event that polluted the early Solar System to be $>$28 Myr based on the \iso{135}Cs/\iso{133}Cs ratio, consistent with the elapsed times derived from \iso{107}Pd and \iso{182}Hf. The abundance sum of \iso{135}Ba and \iso{135}Cs is found to increase, resulting in a smaller $r$-process contribution of \iso{135}Ba to the Solar System.
The remaining uncertainty of the \iso{134}Cs stellar decay rate mainly comes from the tentative spin-parity assignment of some low-lying states. The spin-parity assignments are essential to match shell model levels with experimental levels, and incorrect assignments would yield incorrect matching. Experimental investigation of the spin-parity structure of \iso{134}Cs is therefore key to a precise understanding of multiple astrophysical nucleosynthesis sites in this mass region.

\begin{acknowledgments}
This project is supported by the Strategic Priority Research Program of Chinese Academy of Sciences, Grant No. XDB34020000, the national key research and development program (MOST 2016YFA0400501), the ERC Consolidator Grant (Hungary) funding scheme (Project RADIOSTAR, G.A. n. 724560), the ISSI-Beijing project "Radioactive Nuclei in the Cosmos and in the Solar System" and the Chinese and Hungarian Academy of Sciences visitor exchange program. C.Q. was supported by the Swedish Research Council (VR) under grant Nos. 621-2012-3805, and 621-2013-4323 and the G\"{o}ran Gustafsson foundation. The calculations were performed on resources provided by the Swedish National Infrastructure for Computing (SNIC) at PDC at KTH, Stockholm. A.I.K. was supported by the Australian Research Council Centre of Excellence for All Sky Astrophysics in 3 Dimensions (ASTRO 3D), through project number CE170100013. X.T. acknowledges support from the National Natural Science Foundation of China under Grant No. 11021504, 11321064, 11475228 and 11490564, 100 talents Program of the Chinese Academy of Sciences. X.T. thanks Hayden Campbell for proofreading.
\end{acknowledgments}

\bibliography{references}

%\begin{thebibliography}{}
%\bibitem[Howard et al.(1991)]{ba_p_woosley} W. Howard, B. Meyer, and S. Woosley,\ 1991, \apj, 373, L5
%\bibitem{apj1998} S.Gallino \textit{et al}, Astophys. J.
%\textbf{497}, 388(1998)
%\end{thebibliography}

\end{document}